\documentclass[sigconf]{acmart-me}
\usepackage{booktabs} % For formal tables
\usepackage{url}
\usepackage{color}
\usepackage{enumitem}
\usepackage{graphicx}
\usepackage{subcaption}
\usepackage[strings]{underscore}
\usepackage{multirow}
\setcopyright{rightsretained}
\acmDOI{}

\acmISBN{}

\acmConference[MediaEval'20]{Multimedia Evaluation Workshop}{December 14-15 2020}{Online} 
\acmYear{2020}
\copyrightyear{}

\acmPrice{}

\begin{document}
	%\title{TIB\_VA at MediaEval'20: FakeNews - Corona virus and 5G conspiracy}
	\title{TIB's Visual Analytics Group at MediaEval '20: Detecting Fake News on Corona Virus and 5G Conspiracy}
	%\titlenote{Produces the permission block, and
	%  copyright information}
	%\subtitle{Extended Abstract}
	%\subtitlenote{The full version of the author's guide is available as
	% \texttt{acmart.pdf} document}

	%%If you want to use multi-column authors that's fine. Comment out the next lines, and then uncomment below.
	\author{Gullal S. Cheema\textsuperscript{1}, Sherzod Hakimov\textsuperscript{1}, Ralph Ewerth\textsuperscript{1,2}}
	\affiliation{\textsuperscript{1}TIB -- Leibniz Information Centre for Science and Technology, Hannover, Germany\\ \textsuperscript{2}L3S Research Center, Leibniz University Hannover, Germany}
	\email{{gullal.cheema, sherzod.hakimov, ralph.ewerth}@tib.eu}
	
	%%\authornote{The secretary disavows any knowledge of this author's actions.}
	%\affiliation{Institute for Clarity in Documentation, Ohio, USA}
	%\email{webmaster@marysville-ohio.com}
	%
	%\author{Lars Th{\o}rv{\"a}ld}
	%%\authornote{This author is the
	%%  one who did all the really hard work.}
	%\affiliation{The Th{\o}rv{\"a}ld Group, Iceland}
	%\email{larst@affiliation.org}
	%
	%\author{Lawrence P. Leipuner}
	%\affiliation{Brookhaven Labs, France}
	%\email{lleipuner@researchlabs.org}
	%
	%\author{Sean Fogarty}
	%\affiliation{A Research Institute, Germany}
	%\email{fogartys@amesres.org}
	%\affiliation{Palmer Research Laboratories, Texas, USA}
	%\affiliation{The Th{\o}rv{\"a}ld Group, Iceland}
	%\email{jsmith@affiliation.org}
	
	\renewcommand{\shortauthors}{G.S. Cheema et al.}
	\renewcommand{\shorttitle}{FakeNews: Corona virus and 5G conspiracy}
	
	\begin{abstract}
		Fake news on social media has become a hot topic of research as it negatively impacts the discourse of real news in the public. Specifically, the ongoing \emph{COVID-19} pandemic has seen a rise of inaccurate and misleading information due to the surrounding controversies and unknown details at the beginning of the pandemic. The \emph{FakeNews} task at MediaEval 2020 tackles this problem by creating a challenge to automatically detect tweets containing misinformation based on text and structure from Twitter follower network. In this paper, we present a simple approach that uses BERT embeddings and a shallow neural network for classifying tweets using only text, and discuss our findings and limitations of the approach in text-based misinformation detection.
	\end{abstract}
	
	\maketitle
	
	\section{Introduction and Related Work}
	\label{sec:intro}
	The \emph{FakeNews} task~\cite{pogorelov2020fakenews}\footnote{\url{https://multimediaeval.github.io/editions/2020/tasks/fakenews/}} focuses on automatically predicting whether a tweet consists of misinformation (conspiracy) over the use of two concepts \emph{COVID-19} and \emph{5G} network. The dataset also consists of other conspiracy tweets that are either over some other concepts or accidentally contain the two buzzwords. The challenge requires the participants to mainly develop text or structure based detection models to automatically detect conspiracy tweets. 
	
	In the last five years, social media fake news detection has attracted a lot of research interest in academia and industry. Consequently, the problem has been approached from different perspectives including stance detection~\cite{riedel2017simple,hanselowski2018retrospective}, claim detection and verification~\cite{elsayed2019overview,barron2020overview}, sentiment analysis~\cite{cui2019same,ajao2019sentiment}, etc. To learn a model from text, recently different variants of neural networks have been used for fake news detection. Convolutional Neural Networks (CNN) in general have been extensively used with word embeddings in several works~\cite{yang2018ti,fang2019self,kaliyar2020fndnet} for social media fake news detection. Recently, Ajao \textit{et. al.}~\cite{ajao2018fake} proposed a hybrid model with a Convolutional Neural Network (CNN) and a Recurrent Neural Network (RNN) to identify fake news on Twitter. In similar CLEF challenges~\cite{elsayed2019overview,barron2020overview} over the years, the problem of claim detection in tweets has been tackled with a combination of rich set of features and different kinds of classifiers like SVM~\cite{zuo2018hybrid}, gradient boosting~\cite{yasser2018bigir} and sequential neural networks~\cite{hansen2018copenhagen}. However, several works~\cite{clef-checkthat-williams:2020,cheema2020checksquare} recently have moved from using word2vec~\cite{mikolov2013distributed} type embeddings to rich contextual deep transformer \emph{BERT} (Bidirectional Encoder Representations from Transformers)~\cite{devlin2018bert} like embeddings.
	
	\section{Approach}
	\label{sec:approach}
	We approach the problem from only the textual perspective and rely on training a shallow neural network over contextual word embeddings. Our submitted models use the recently proposed \emph{BERT-large} based model pre-trained~\cite{muller2020covid} on a large corpus of \emph{COVID} Twitter data. This essentially improves the performance by 3-4\% in comparison to vanilla \emph{BERT} (V-BERT) since the embeddings are better aligned (with regard to \emph{COVID}) for the task at hand. We also experiment with additional features like sentiment, subjectivity and lexical features that have been shown to improve performance in similar tasks \cite{elsayed2019overview,barron2020overview}. We observed no improvements using combination of these features and excluded them in this paper.\footnote{Source code: \url{https://github.com/cleopatra-itn/TIB_VA_MediaEval_FakeNews}}
	
	\noindent\textbf{Text Pre-processing}
	For vanilla \emph{BERT-large}, we use Baziotis \textit{et. al.}'s~\cite{baziotis-pelekis-doulkeridis:2017:SemEval2} tool to apply the following normalization steps: tokenization, lower-casing, removal of punctuation, spell correction, normalize \textit{hashtags, all-caps, censored, elongated and repeated} words, and remove terms like \textit{URL, email, phone, user mentions}. For \emph{COVID} Twitter BERT~\cite{muller2020covid}, we follow their pre-processing which normalizes text, and additionally replaces \textit{user mentions, emails, URLs} with special keywords.
	
	\noindent\textbf{Contextual Feature Extraction}
	To get one embedding per tweet, we follow the observations made by \citet{devlin2018bert} that different layers of \emph{BERT} capture different kinds of information, so an appropriate pooling strategy should be applied depending on the task. The paper also suggests that the last four hidden layers of the network are good for transfer learning tasks and thus we experiment with 4 different combinations, i.e., concatenate last 4 hidden layers (4-CAT), the average of last 4 hidden layers (4-SUM), last hidden layer (LAST), and $2^{nd}$ last hidden layer (2-LAST). We normalize the final embedding so that $l2$ norm of the vector is 1.
	
	\noindent\textbf{Shallow Neural Network}
	We use the extracted and pooled \emph{BERT} embeddings and train a two-layer neural network. Before passing the features through the first layer, we apply a squeeze and excitation (SE) operation \cite{hu2018squeeze} that enhances the representation to learn a better model. The SE operation has been shown to improve feature representations and performance in CNNs. Then, the embedding is projected down to 128 dimensions, which is followed by batch normalization and ReLU operation to introduce non-linearity. The $2^{nd}$ layer is a linear classification layer that produces a softmax probability for each class. Dropout with rate of 0.2 and 0.5 is applied after SE operation and first layer to avoid over-fitting.
	
	\noindent\textbf{Final Prediction on Test Set}
	We train five models on 5-fold splits and take the majority label as the predicted label for the tweet. As described in the challenge, the 3-class submissions can have an additional \textit{cannot-determine} class. We assign a tweet with this label if the softmax probability is less than 0.4, which signifies that the model is not confident enough.
% 	As one of the classes can be \textit{cannot-determine}, we assign a tweet with this label if the softmax probability is less than 0.4, which signifies that the model is not confident enough. 

	\section{Experimental Results}
	\label{sec:results}
	The \emph{FakeNews} development set consists of 5,999 extracted tweets over three classes: \emph{5G} and \emph{COVID-19} conspiracy (1,128 samples), Non-conspiracy (4,173 samples) and other conspiracy (698 samples). We generate five-fold stratified training and validation splits in the ratio of 80:20, so that the distribution of classes remains the same in the training and validation sets. The official test data originally consisted of 3,230 tweets, out of which 308 are not valid or non-existent at the time of evaluation since participants were required to crawl tweets on their own. Table~\ref{tab:val} shows the performance of different models on validation sets, while Table~\ref{tab:test} shows the evaluation of our submitted models on the official test data. All the runs for the test data use \emph{COVID} Twitter BERT (C-BERT) extracted features. Official metric is Matthews correlation coefficient (MCC). For validation sets, we provide both average accuracy (ACC) and MCC scores. In Table~\ref{tab:val}, we also show the result of fine-tuning (FT) the last two and four layers of 2 BERT variants with a linear classification layer on top of BERT's \emph{CLS} embedding. Although the average performance of finetuning 4 layers is marginally better than the fixed average word embeddings, the highest in two of the splits is better in the fixed embedding plus neural network.
	
\bgroup
\setlength{\tabcolsep}{1.1pt} % Default value: 6pt
\renewcommand{\arraystretch}{1.1} % Default value: 1
	\begin{table}[b]%!ht]
		\centering
		\caption{Model Evaluation of Validation Set.}
		\begin{tabular}{|l|c|c|c|c|}
			\hline
			\textbf{Model}             & \textbf{Pooling}               & \textbf{Classes} & \textbf{ACC}                   & \textbf{MCC}                   \\ \hline
			V-BERT (FT)$^{2}$ & \emph{CLS}             & 3       &      0.7616 $\pm$ 0.0125                 &     0.4521 $\pm$ 0.0347                  \\ \hline
			V-BERT (FT)$^{4}$ & \emph{CLS}             & 3       &      0.7656 $\pm$ 0.0094                &     0.4611 $\pm$ 0.0266                  \\ \hline
			C-BERT (FT)$^{2}$   & \emph{CLS}             & 3       &  0.8131 $\pm$ 0.0063                 &     0.5837 $\pm$ 0.0107                     \\ \hline
			C-BERT (FT)$^{4}$   & \emph{CLS}             & 3       &      \textbf{0.8171 $\pm$ 0.0091}                 &      \textbf{0.5952 $\pm$ 0.0192}                 \\ \hline
			\multirow{4}{*}{C-BERT + NN}   & 4-SUM                 & 3       &       0.8138 $\pm$ 0.0112                &       0.5793 $\pm$ 0.0272                \\ \cline{2-5}
			& 4-CAT                 & 3       &         \textbf{0.8163 $\pm$ 0.0112}              &      \textbf{ 0.5841 $\pm$ 0.0264}                \\ \cline{2-5}
			& 2-LAST                 & 3       &  0.813 $\pm$ 0.0073              &       0.5761 $\pm$ 0.0181  \\ \cline{2-5}
			& LAST                  & 3       &  0.8083 $\pm$ 0.0085              &       0.5662 $\pm$ 0.0211  \\ \hline
			\multirow{4}{*}{C-BERT + NN}   & 4-SUM                 & 2       &    0.8881 $\pm$ 0.0122              &       0.6173 $\pm$ 0.0401                       \\ \cline{2-5}
			& 4-CAT                 & 2       &    \textbf{0.8901 $\pm$ 0.0087}              &       \textbf{0.6236 $\pm$ 0.0330}                    \\ \cline{2-5}
			& 2-LAST                & 2       &  0.8886 $\pm$ 0.0103              &       0.6184 $\pm$ 0.0357 \\ \cline{2-5}
			& LAST                  & 2       &  0.8838 $\pm$ 0.0058              &       0.5976 $\pm$ 0.0218  \\ \hline
		\end{tabular}
		\label{tab:val}
	\end{table}
\egroup
	
\bgroup
\setlength{\tabcolsep}{5pt} % Default value: 6pt
\renewcommand{\arraystretch}{1.0} % Default value: 1
	\begin{table}[!ht]
		\centering
		\caption{Official Test Data Results. 3-class submissions include an additional \textit{cannot-determine} (CD) class when the model confidence is lower than 0.4.}
		\begin{tabular}{|l|c|c|}
			\hline
			\textbf{Submission} & \textbf{Classes} & \textbf{MCC}    \\ \hline
			Run 1 (4-SUM)     & 3+CD       & 0.5717 \\ \hline
			Run 2 (4-CAT)     & 3+CD       & 0.5773 \\ \hline
			Run 3 (4-SUM)     & 2       & 0.5986  \\ \hline
			Run 4 (4-CAT)     & 2       & 0.6083 \\ \hline
		\end{tabular}
		\label{tab:test}
	\end{table}
\egroup
	
\section{Discussion}
\label{sec:discussion}
\begin{figure}[t]
    \vspace{-0.3cm}
	\centering
	\begin{subfigure}[b]{0.23\textwidth}
		\centering
		\includegraphics[width=\textwidth]{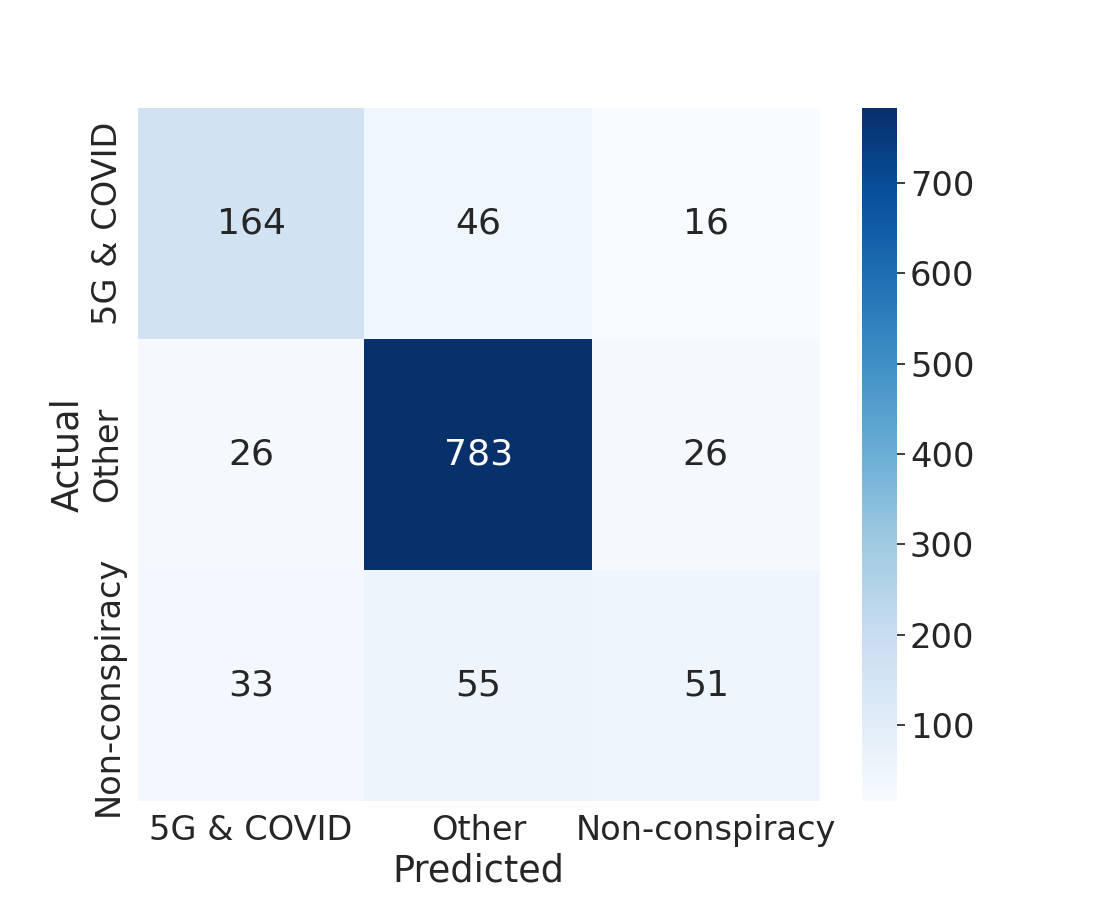}
		\caption{3-class}
		\label{fig:3cls}
	\end{subfigure}
	\begin{subfigure}[b]{0.23\textwidth}
		\centering
		\includegraphics[width=\textwidth]{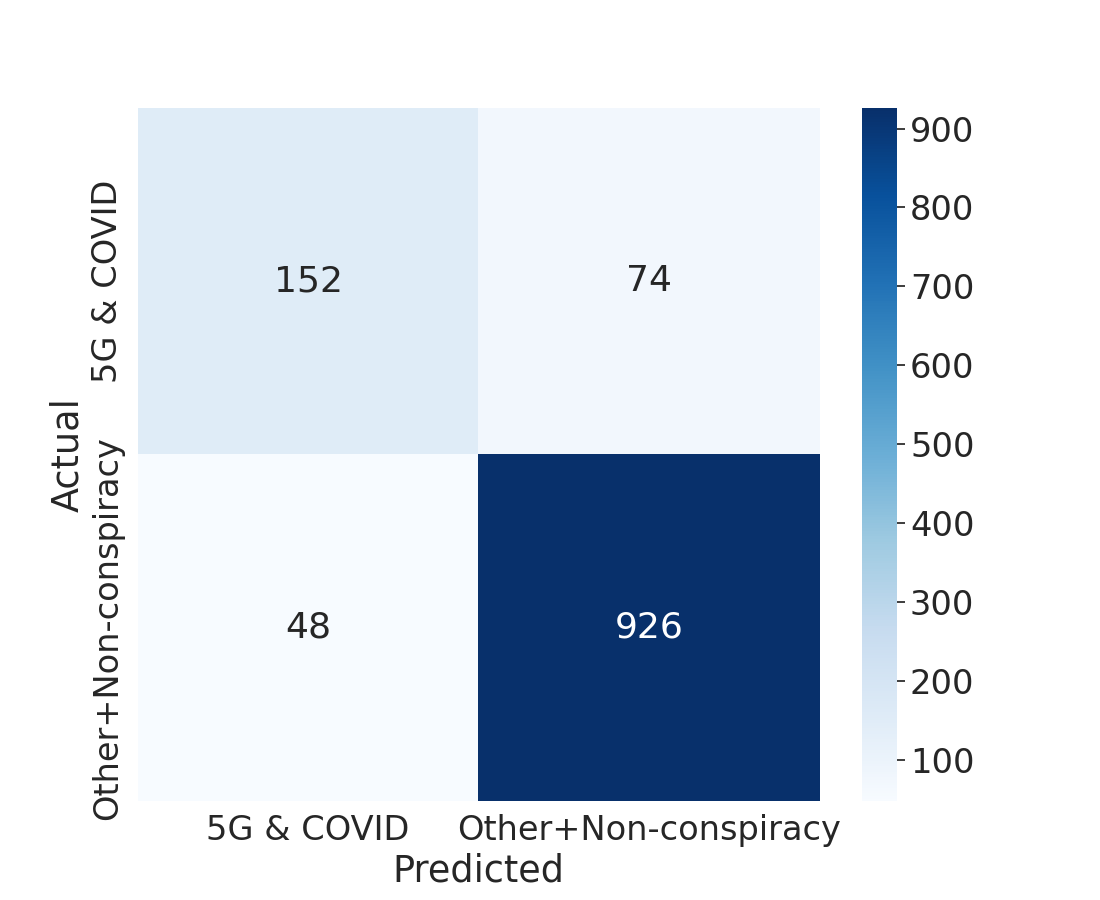}
		\caption{2-class}
		\label{fig:2cls}
	\end{subfigure}
	\caption{Confusion matrices for the different settings}
	\label{fig:confmat}
	\vspace{-0.5cm}
\end{figure}

Our findings and observations from the FakeNews task can be summarized as follows:
\begin{itemize}
	\item Vanilla \emph{BERT} clearly has a wider domain gap to perform well on this task, as the concepts and keywords related to \emph{COVID} are fairly recent. The \emph{COVID} Twitter \emph{BERT} outperforms in our finetuning experiment as well as with a shallow neural network on the extracted embeddings.
	\item The pooling operation and the number of last layers to obtain a sentence embedding does make a difference, as only using the last layer (or $2^{nd}$ last) performs marginally lower across the metrics. An even better embedding could be a sentence embedding extracted from a sentence-transformer \cite{reimers2019sentence}, but only if it is pretrained on a \emph{COVID} Twitter corpus to narrow down the domain and knowledge gap.
	\item Although two-class prediction performance has higher metric scores, merging the other-conspiracy and \\
	non-conspiracy tweets decreases the true positives (see Figure~\ref{fig:confmat}) for the conspiracy class. This could be because the model is able to learn and detect the conspiracy aspect in tweets, and merging the other two categories negatively impacts the learning.
	\item In similar social media challenges, pre-processing text also plays a significant role. Therefore, we experimented with replacing different keywords like \textit{corona, sars cov2, wuhan virus, ncov, korona, koronavirus} with \textit{coronavirus} or \textit{covid}, and similarly \textit{five g, fiveg, 5 g} with \textit{5g}. Unfortunately, doing so degraded the performance in some splits and was not a part of our submission model.
\end{itemize}

\section{Conclusion}
In this paper, we have presented our solution for the FakeNews detection task of MediaEval 2020. The described solution is based on extracting embeddings from transformer models and training shallow neural networks. We compared the two transformer models and observed that \emph{BERT} transformer pre-trained on \emph{COVID} tweets performs better than vanilla version. Pooling operations such as concatenation or averaging of embeddings of the last hidden layers also play an important role as shown by experimental evaluation. In future work, we will focus on the integration of additional contextual information that is presented via external links along with data from other modalities such as images.

\section*{Acknowledgements}
This project has received funding from the European Union’s Horizon 2020 research and innovation programme under the Marie Skłodowska-Curie grant agreement no 812997 (CLEOPATRA ITN).

\bibliographystyle{ACM-Reference-Format}
\def\bibfont{\small} % comment this line for a smaller fontsize
\bibliography{main} 


\begin{thebibliography}{00}

%%% the Web address.

\ifx \showCODEN    \undefined \def \showCODEN     #1{\unskip}     \fi
\ifx \showDOI      \undefined \def \showDOI       #1{#1}\fi
\ifx \showISBNx    \undefined \def \showISBNx     #1{\unskip}     \fi
\ifx \showISBNxiii \undefined \def \showISBNxiii  #1{\unskip}     \fi
\ifx \showISSN     \undefined \def \showISSN      #1{\unskip}     \fi
\ifx \showLCCN     \undefined \def \showLCCN      #1{\unskip}     \fi
\ifx \shownote     \undefined \def \shownote      #1{#1}          \fi
\ifx \showarticletitle \undefined \def \showarticletitle #1{#1}   \fi
\ifx \showURL      \undefined \def \showURL       {\relax}        \fi
% invisible to TeX
\providecommand\bibfield[2]{#2}
\providecommand\bibinfo[2]{#2}
\providecommand\natexlab[1]{#1}
\providecommand\showeprint[2][]{arXiv:#2}

\bibitem[\protect\citeauthoryear{Ajao, Bhowmik, and Zargari}{Ajao
  et~al\mbox{.}}{2018}]%
        {ajao2018fake}
\bibfield{author}{\bibinfo{person}{Oluwaseun Ajao}, \bibinfo{person}{Deepayan
  Bhowmik}, {and} \bibinfo{person}{Shahrzad Zargari}.}
  \bibinfo{year}{2018}\natexlab{}.
\newblock \showarticletitle{Fake news identification on twitter with hybrid cnn
  and rnn models}. In \bibinfo{booktitle}{{\em Proceedings of the 9th
  international conference on social media and society}}.
  \bibinfo{pages}{226--230}.
\newblock


\bibitem[\protect\citeauthoryear{Ajao, Bhowmik, and Zargari}{Ajao
  et~al\mbox{.}}{2019}]%
        {ajao2019sentiment}
\bibfield{author}{\bibinfo{person}{Oluwaseun Ajao}, \bibinfo{person}{Deepayan
  Bhowmik}, {and} \bibinfo{person}{Shahrzad Zargari}.}
  \bibinfo{year}{2019}\natexlab{}.
\newblock \showarticletitle{Sentiment aware fake news detection on online
  social networks}. In \bibinfo{booktitle}{{\em ICASSP 2019-2019 IEEE
  International Conference on Acoustics, Speech and Signal Processing
  (ICASSP)}}. IEEE, \bibinfo{pages}{2507--2511}.
\newblock


\bibitem[\protect\citeauthoryear{Barr{\'o}n-Cedeno, Elsayed, Nakov,
  Da~San~Martino, Hasanain, Suwaileh, Haouari, Babulkov, Hamdan, Nikolov,
  et~al\mbox{.}}{Barr{\'o}n-Cedeno et~al\mbox{.}}{2020}]%
        {barron2020overview}
\bibfield{author}{\bibinfo{person}{Alberto Barr{\'o}n-Cedeno},
  \bibinfo{person}{Tamer Elsayed}, \bibinfo{person}{Preslav Nakov},
  \bibinfo{person}{Giovanni Da~San~Martino}, \bibinfo{person}{Maram Hasanain},
  \bibinfo{person}{Reem Suwaileh}, \bibinfo{person}{Fatima Haouari},
  \bibinfo{person}{Nikolay Babulkov}, \bibinfo{person}{Bayan Hamdan},
  \bibinfo{person}{Alex Nikolov}, {and} \bibinfo{person}{others}.}
  \bibinfo{year}{2020}\natexlab{}.
\newblock \showarticletitle{Overview of CheckThat! 2020: Automatic
  identification and verification of claims in social media}. In
  \bibinfo{booktitle}{{\em International Conference of the Cross-Language
  Evaluation Forum for European Languages}}. Springer,
  \bibinfo{pages}{215--236}.
\newblock


\bibitem[\protect\citeauthoryear{Baziotis, Pelekis, and Doulkeridis}{Baziotis
  et~al\mbox{.}}{2017}]%
        {baziotis-pelekis-doulkeridis:2017:SemEval2}
\bibfield{author}{\bibinfo{person}{Christos Baziotis}, \bibinfo{person}{Nikos
  Pelekis}, {and} \bibinfo{person}{Christos Doulkeridis}.}
  \bibinfo{year}{2017}\natexlab{}.
\newblock \showarticletitle{DataStories at SemEval-2017 Task 4: Deep LSTM with
  Attention for Message-level and Topic-based Sentiment Analysis}. In
  \bibinfo{booktitle}{{\em Proceedings of the 11th International Workshop on
  Semantic Evaluation (SemEval-2017)}}. \bibinfo{publisher}{Association for
  Computational Linguistics}, \bibinfo{address}{Vancouver, Canada},
  \bibinfo{pages}{747--754}.
\newblock


\bibitem[\protect\citeauthoryear{Cheema, Hakimov, and Ewerth}{Cheema
  et~al\mbox{.}}{2020}]%
        {cheema2020checksquare}
\bibfield{author}{\bibinfo{person}{Gullal~S. Cheema}, \bibinfo{person}{Sherzod
  Hakimov}, {and} \bibinfo{person}{Ralph Ewerth}.}
  \bibinfo{year}{2020}\natexlab{}.
\newblock \showarticletitle{Check{\_}square at CheckThat! 2020 Claim Detection
  in Social Media via Fusion of Transformer and Syntactic Features}. In
  \bibinfo{booktitle}{{\em Working Notes of {CLEF} 2020 - Conference and Labs
  of the Evaluation Forum, Thessaloniki, Greece, September 22-25, 2020}} {\em
  (\bibinfo{series}{{CEUR} Workshop Proceedings})},
  \bibfield{editor}{\bibinfo{person}{Linda Cappellato},
  \bibinfo{person}{Carsten Eickhoff}, \bibinfo{person}{Nicola Ferro}, {and}
  \bibinfo{person}{Aur{\'{e}}lie N{\'{e}}v{\'{e}}ol}} (Eds.),
  Vol.~\bibinfo{volume}{2696}. \bibinfo{publisher}{CEUR-WS.org}.
\newblock
\showURL{%
\url{http://ceur-ws.org/Vol-2696/paper\_216.pdf}}


\bibitem[\protect\citeauthoryear{Cui, Wang, and Lee}{Cui et~al\mbox{.}}{2019}]%
        {cui2019same}
\bibfield{author}{\bibinfo{person}{Limeng Cui}, \bibinfo{person}{Suhang Wang},
  {and} \bibinfo{person}{Dongwon Lee}.} \bibinfo{year}{2019}\natexlab{}.
\newblock \showarticletitle{SAME: sentiment-aware multi-modal embedding for
  detecting fake news}. In \bibinfo{booktitle}{{\em Proceedings of the 2019
  IEEE/ACM International Conference on Advances in Social Networks Analysis and
  Mining}}. \bibinfo{pages}{41--48}.
\newblock


\bibitem[\protect\citeauthoryear{Devlin, Chang, Lee, and Toutanova}{Devlin
  et~al\mbox{.}}{2019}]%
        {devlin2018bert}
\bibfield{author}{\bibinfo{person}{Jacob Devlin}, \bibinfo{person}{Ming{-}Wei
  Chang}, \bibinfo{person}{Kenton Lee}, {and} \bibinfo{person}{Kristina
  Toutanova}.} \bibinfo{year}{2019}\natexlab{}.
\newblock \showarticletitle{{BERT:} Pre-training of Deep Bidirectional
  Transformers for Language Understanding}. In \bibinfo{booktitle}{{\em
  Proceedings of the 2019 Conference of the North American Chapter of the
  Association for Computational Linguistics}}. \bibinfo{publisher}{Association
  for Computational Linguistics}, \bibinfo{pages}{4171--4186}.
\newblock
\showDOI{%
\url{https://doi.org/10.18653/v1/n19-1423}}


\bibitem[\protect\citeauthoryear{Elsayed, Nakov, Barr{\'o}n-Cedeno, Hasanain,
  Suwaileh, Da~San~Martino, and Atanasova}{Elsayed et~al\mbox{.}}{2019}]%
        {elsayed2019overview}
\bibfield{author}{\bibinfo{person}{Tamer Elsayed}, \bibinfo{person}{Preslav
  Nakov}, \bibinfo{person}{Alberto Barr{\'o}n-Cedeno}, \bibinfo{person}{Maram
  Hasanain}, \bibinfo{person}{Reem Suwaileh}, \bibinfo{person}{Giovanni
  Da~San~Martino}, {and} \bibinfo{person}{Pepa Atanasova}.}
  \bibinfo{year}{2019}\natexlab{}.
\newblock \showarticletitle{Overview of the CLEF-2019 CheckThat! Lab: automatic
  identification and verification of claims}. In \bibinfo{booktitle}{{\em
  International Conference of the Cross-Language Evaluation Forum for European
  Languages}}. Springer, \bibinfo{pages}{301--321}.
\newblock


\bibitem[\protect\citeauthoryear{Fang, Gao, Huang, Peng, and Wu}{Fang
  et~al\mbox{.}}{2019}]%
        {fang2019self}
\bibfield{author}{\bibinfo{person}{Yong Fang}, \bibinfo{person}{Jian Gao},
  \bibinfo{person}{Cheng Huang}, \bibinfo{person}{Hua Peng}, {and}
  \bibinfo{person}{Runpu Wu}.} \bibinfo{year}{2019}\natexlab{}.
\newblock \showarticletitle{Self multi-head attention-based convolutional
  neural networks for fake news detection}.
\newblock \bibinfo{journal}{{\em PloS one\/}} \bibinfo{volume}{14},
  \bibinfo{number}{9} (\bibinfo{year}{2019}), \bibinfo{pages}{e0222713}.
\newblock


\bibitem[\protect\citeauthoryear{Hanselowski, PVS, Schiller, Caspelherr,
  Chaudhuri, Meyer, and Gurevych}{Hanselowski et~al\mbox{.}}{2018}]%
        {hanselowski2018retrospective}
\bibfield{author}{\bibinfo{person}{Andreas Hanselowski},
  \bibinfo{person}{Avinesh PVS}, \bibinfo{person}{Benjamin Schiller},
  \bibinfo{person}{Felix Caspelherr}, \bibinfo{person}{Debanjan Chaudhuri},
  \bibinfo{person}{Christian~M Meyer}, {and} \bibinfo{person}{Iryna Gurevych}.}
  \bibinfo{year}{2018}\natexlab{}.
\newblock \showarticletitle{A retrospective analysis of the fake news challenge
  stance detection task}.
\newblock \bibinfo{journal}{{\em arXiv preprint arXiv:1806.05180\/}}
  (\bibinfo{year}{2018}).
\newblock


\bibitem[\protect\citeauthoryear{Hansen, Hansen, Simonsen, and Lioma}{Hansen
  et~al\mbox{.}}{2018}]%
        {hansen2018copenhagen}
\bibfield{author}{\bibinfo{person}{Casper Hansen}, \bibinfo{person}{Christian
  Hansen}, \bibinfo{person}{Jakob~Grue Simonsen}, {and}
  \bibinfo{person}{Christina Lioma}.} \bibinfo{year}{2018}\natexlab{}.
\newblock \showarticletitle{The Copenhagen Team Participation in the
  Check-Worthiness Task of the Competition of Automatic Identification and
  Verification of Claims in Political Debates of the CLEF-2018 CheckThat!
  Lab.}. In \bibinfo{booktitle}{{\em CLEF (Working Notes)}}.
\newblock


\bibitem[\protect\citeauthoryear{Hu, Shen, and Sun}{Hu et~al\mbox{.}}{2018}]%
        {hu2018squeeze}
\bibfield{author}{\bibinfo{person}{Jie Hu}, \bibinfo{person}{Li Shen}, {and}
  \bibinfo{person}{Gang Sun}.} \bibinfo{year}{2018}\natexlab{}.
\newblock \showarticletitle{Squeeze-and-excitation networks}. In
  \bibinfo{booktitle}{{\em Proceedings of the IEEE conference on computer
  vision and pattern recognition}}. \bibinfo{pages}{7132--7141}.
\newblock


\bibitem[\protect\citeauthoryear{Kaliyar, Goswami, Narang, and Sinha}{Kaliyar
  et~al\mbox{.}}{2020}]%
        {kaliyar2020fndnet}
\bibfield{author}{\bibinfo{person}{Rohit~Kumar Kaliyar},
  \bibinfo{person}{Anurag Goswami}, \bibinfo{person}{Pratik Narang}, {and}
  \bibinfo{person}{Soumendu Sinha}.} \bibinfo{year}{2020}\natexlab{}.
\newblock \showarticletitle{FNDNet--A deep convolutional neural network for
  fake news detection}.
\newblock \bibinfo{journal}{{\em Cognitive Systems Research\/}}
  \bibinfo{volume}{61} (\bibinfo{year}{2020}), \bibinfo{pages}{32--44}.
\newblock


\bibitem[\protect\citeauthoryear{Mikolov, Sutskever, Chen, Corrado, and
  Dean}{Mikolov et~al\mbox{.}}{2013}]%
        {mikolov2013distributed}
\bibfield{author}{\bibinfo{person}{Tomas Mikolov}, \bibinfo{person}{Ilya
  Sutskever}, \bibinfo{person}{Kai Chen}, \bibinfo{person}{Greg~S Corrado},
  {and} \bibinfo{person}{Jeff Dean}.} \bibinfo{year}{2013}\natexlab{}.
\newblock \showarticletitle{Distributed representations of words and phrases
  and their compositionality}.
\newblock \bibinfo{journal}{{\em Advances in neural information processing
  systems\/}}  \bibinfo{volume}{26} (\bibinfo{year}{2013}),
  \bibinfo{pages}{3111--3119}.
\newblock


\bibitem[\protect\citeauthoryear{M{\"u}ller, Salath{\'e}, and
  Kummervold}{M{\"u}ller et~al\mbox{.}}{2020}]%
        {muller2020covid}
\bibfield{author}{\bibinfo{person}{Martin M{\"u}ller}, \bibinfo{person}{Marcel
  Salath{\'e}}, {and} \bibinfo{person}{Per~E Kummervold}.}
  \bibinfo{year}{2020}\natexlab{}.
\newblock \showarticletitle{COVID-Twitter-BERT: A Natural Language Processing
  Model to Analyse COVID-19 Content on Twitter}.
\newblock \bibinfo{journal}{{\em arXiv preprint arXiv:2005.07503\/}}
  (\bibinfo{year}{2020}).
\newblock


\bibitem[\protect\citeauthoryear{Pogorelov, Schroeder, Burchard, Moe, Brenner,
  Filkukova, and Langguth}{Pogorelov et~al\mbox{.}}{2020}]%
        {pogorelov2020fakenews}
\bibfield{author}{\bibinfo{person}{Konstantin Pogorelov},
  \bibinfo{person}{Daniel~Thilo Schroeder}, \bibinfo{person}{Luk Burchard},
  \bibinfo{person}{Johannes Moe}, \bibinfo{person}{Stefan Brenner},
  \bibinfo{person}{Petra Filkukova}, {and} \bibinfo{person}{Johannes
  Langguth}.} \bibinfo{year}{2020}\natexlab{}.
\newblock \showarticletitle{FakeNews: Corona Virus and 5G Conspiracy Task at
  MediaEval 2020}. In \bibinfo{booktitle}{{\em MediaEval 2020 Workshop}}.
\newblock


\bibitem[\protect\citeauthoryear{Reimers and Gurevych}{Reimers and
  Gurevych}{2019}]%
        {reimers2019sentence}
\bibfield{author}{\bibinfo{person}{Nils Reimers} {and} \bibinfo{person}{Iryna
  Gurevych}.} \bibinfo{year}{2019}\natexlab{}.
\newblock \showarticletitle{Sentence-BERT: Sentence Embeddings using Siamese
  BERT-Networks}. In \bibinfo{booktitle}{{\em Proceedings of the 2019
  Conference on Empirical Methods in Natural Language Processing and the 9th
  International Joint Conference on Natural Language Processing
  (EMNLP-IJCNLP)}}. \bibinfo{pages}{3973--3983}.
\newblock


\bibitem[\protect\citeauthoryear{Riedel, Augenstein, Spithourakis, and
  Riedel}{Riedel et~al\mbox{.}}{2017}]%
        {riedel2017simple}
\bibfield{author}{\bibinfo{person}{Benjamin Riedel}, \bibinfo{person}{Isabelle
  Augenstein}, \bibinfo{person}{Georgios~P Spithourakis}, {and}
  \bibinfo{person}{Sebastian Riedel}.} \bibinfo{year}{2017}\natexlab{}.
\newblock \showarticletitle{A simple but tough-to-beat baseline for the Fake
  News Challenge stance detection task}.
\newblock \bibinfo{journal}{{\em arXiv preprint arXiv:1707.03264\/}}
  (\bibinfo{year}{2017}).
\newblock


\bibitem[\protect\citeauthoryear{Williams, Rodrigues, and Novak}{Williams
  et~al\mbox{.}}{2020}]%
        {clef-checkthat-williams:2020}
\bibfield{author}{\bibinfo{person}{Evan Williams}, \bibinfo{person}{Paul
  Rodrigues}, {and} \bibinfo{person}{Valerie Novak}.}
  \bibinfo{year}{2020}\natexlab{}.
\newblock \showarticletitle{Accenture at CheckThat!~2020: If you say so:
  Post-hoc fact-checking of Claims using Transformer-based Models}. In
  \bibinfo{booktitle}{{\em Working Notes of {CLEF} 2020 - Conference and Labs
  of the Evaluation Forum, Thessaloniki, Greece, September 22-25, 2020}} {\em
  (\bibinfo{series}{{CEUR} Workshop Proceedings})},
  \bibfield{editor}{\bibinfo{person}{Linda Cappellato},
  \bibinfo{person}{Carsten Eickhoff}, \bibinfo{person}{Nicola Ferro}, {and}
  \bibinfo{person}{Aur{\'{e}}lie N{\'{e}}v{\'{e}}ol}} (Eds.),
  Vol.~\bibinfo{volume}{2696}. \bibinfo{publisher}{CEUR-WS.org}.
\newblock
\showURL{%
\url{http://ceur-ws.org/Vol-2696/paper\_226.pdf}}


\bibitem[\protect\citeauthoryear{Yang, Zheng, Zhang, Cui, Li, and Yu}{Yang
  et~al\mbox{.}}{2018}]%
        {yang2018ti}
\bibfield{author}{\bibinfo{person}{Yang Yang}, \bibinfo{person}{Lei Zheng},
  \bibinfo{person}{Jiawei Zhang}, \bibinfo{person}{Qingcai Cui},
  \bibinfo{person}{Zhoujun Li}, {and} \bibinfo{person}{Philip~S Yu}.}
  \bibinfo{year}{2018}\natexlab{}.
\newblock \showarticletitle{TI-CNN: Convolutional neural networks for fake news
  detection}.
\newblock \bibinfo{journal}{{\em arXiv preprint arXiv:1806.00749\/}}
  (\bibinfo{year}{2018}).
\newblock


\bibitem[\protect\citeauthoryear{Yasser, Kutlu, and Elsayed}{Yasser
  et~al\mbox{.}}{2018}]%
        {yasser2018bigir}
\bibfield{author}{\bibinfo{person}{Khaled Yasser}, \bibinfo{person}{Mucahid
  Kutlu}, {and} \bibinfo{person}{Tamer Elsayed}.}
  \bibinfo{year}{2018}\natexlab{}.
\newblock \showarticletitle{bigIR at CLEF 2018: Detection and Verification of
  Check-Worthy Political Claims.}. In \bibinfo{booktitle}{{\em CLEF (Working
  Notes)}}.
\newblock


\bibitem[\protect\citeauthoryear{Zuo, Karakas, and Banerjee}{Zuo
  et~al\mbox{.}}{2018}]%
        {zuo2018hybrid}
\bibfield{author}{\bibinfo{person}{Chaoyuan Zuo}, \bibinfo{person}{Ayla
  Karakas}, {and} \bibinfo{person}{Ritwik Banerjee}.}
  \bibinfo{year}{2018}\natexlab{}.
\newblock \showarticletitle{A Hybrid Recognition System for Check-worthy Claims
  Using Heuristics and Supervised Learning.}. In \bibinfo{booktitle}{{\em CLEF
  (Working Notes)}}.
\newblock


\end{thebibliography}
	
\end{document}